\documentclass[runningheads]{llncs}
\usepackage{bm}
\usepackage{amsfonts,amssymb,amsmath}
\usepackage{graphicx}
\usepackage{xcolor}
\usepackage[colorlinks,linkcolor=violet,citecolor=violet]{hyperref}
\usepackage{url}
\usepackage{epstopdf}
\usepackage{array}
\usepackage{booktabs}
\usepackage{multirow}
\usepackage{subfigure}

\begin{document}

\title{Neural Architecture Search for Gliomas Segmentation on Multimodal Magnetic Resonance Imaging\thanks{
The source code could be found on https://github.com/woodywff/nas3dunet}}

\author{Feifan Wang\inst{1}\orcidID{0000-0002-9525-0656} 
}

\authorrunning{F. Wang et al.}


\institute{School of Life Science and Technology, University of Electronic Science
and Technology of China, Chengdu 611731, China. 
\email{woodywff@aliyun.com}
}
\maketitle              

\begin{abstract}
Past few years have witnessed the artificial intelligence inspired evolution in various medical fields. The diagnosis and treatment of gliomas --- one of the most commonly seen brain tumors with low survival rate --- rely heavily on the computer assisted segmentation process undertaken on the magnetic resonance imaging (MRI) scans. 
Although the encoder-decoder shaped deep learning networks have been the de facto standard style for semantic segmentation tasks in medical imaging analysis, enormous effort is still required to be spent on designing the detailed architecture of the down-sampling and up-sampling blocks.
In this work, we propose a neural architecture search (NAS) based solution to brain tumor segmentation tasks on multimodal volumetric MRI scans. Three sets of candidate operations are composed respectively for three kinds of basic building blocks in which each operation is assigned with a specific probabilistic parameter to be learned. Through alternately updating the weights of operations and the other parameters in the network, the searching mechanism ends up with two optimal structures for the upward and downward blocks. Moreover, the developed solution also integrates normalization and patching strategies tailored for brain MRI processing.  
Extensive comparative experiments on the BraTS 2019 dataset demonstrate that the proposed algorithm not only could relieve the pressure of fabricating block architectures but also possesses competitive feasibility and scalability.
\end{abstract}

\section{Introduction}
The human brain is stable under normal conditions. Nevertheless, this balance would be compromised by the presence of brain tumors, which pathologically are clusters of dysfunctional brain cells~\cite{brain_0}. 
According to their different origins, brain tumors could be classified into the primary brain tumor and the secondary brain tumor. The first one starts from the brain area, while the second one which is also called the metastatic brain tumor is transfered from the other organs in human body. 
From another perspective, brain tumors can also be categorized as malignant or benign. The difference is the malignant tumors are cancerous and likely to be spread to the whole brain while the benign ones are not. 
Gliomas, as one of the most common primary malignant brain tumors, keep attracting researchers' attentions because they could result in more suffering and lost than any other brain tumors~\cite{eighty_percent}. 

Magnetic resonance imaging (MRI) processing, in particular the semantic segmentation telling the tumor tissues apart from the other parts in the brain volume, plays important roles in the diagnosis and treatment of gliomas~\cite{lenting2017glioma}.One vital task for neurosurgeons before the resection surgery is to annotate the tumor regions as precisely as possible, since an ideal brain tumor segmentation could not only preserve enough healthy tissues but also prevent the subsequent tumor recurrence~\cite{kwon2013portr}. During the past few years, scientists from academia and industry have been sparing no effort in exploring computer assisted solutions to give doctors a relief from the laborious and time consuming annotation work.  

The prevail of deep learning has pushed the evolution of semantic segmentation methods. The fully convolution networks (FCN) started the trial of taking advantage of the convolution neural network (CNN) to do the dense image prediction~\cite{fcn}. Intrigued by the similar idea of FCN, many edge-cutting CNNs found their ways in the segmentation tasks, for example the fully convolutional dense nets (FCDN)~\cite{tiramisu} and the DeepLab~\cite{deeplab}. In 2015, the U-Net was firstly brought out aiming at the drosophila cell tracking task in the IEEE international symposium on biomedical imaging (ISBI) challenge~\cite{unet}. Soon after that, it quickly became renowned for the demonstrated effectiveness and efficiency on two dimensional (2D) and three dimensional (3D) medical image datasets~\cite{3dunet}. 

For the gliomas segmentation, U-Net is also one of the most frequently chosen architecture styles. For instance, Kamnitsas et al. made an ensemble of FCN, U-Net and DeepMedic~\cite{kamnitsas2017efficient}, which is a variant of DeepLab for brain lesion segmentations, and won the first place award in the multimodal brain tumor segmentation challenge (BraTS) in 2017~\cite{kamnitsas2017ensembles}. Myronenko, winner of the BraTS 2018, benefited from a combination of the variational auto-encoder (VAE) and the U-Net structure~\cite{nvidia}. Isensee et al. proposed the argument that a well trained U-Net would suffice for the segmentation task, with no need to the extra accessories~\cite{isensee2018no}.
However, even if that is the case, picking up the right building blocks for the down-sampling and up-sampling routes in a U-Net could still be a hard work, considering the diversity of off-the-shelf operation modules may result in a huge variety of candidate architectures. 

As a branch of the automated machine learning (AutoML), NAS specifically focuses on finding out the optimal network structure among the numerous candidate architectures automatically. The blossom of NAS attributes to Zoph and Le who firstly came up with the idea of training a recurrent neural network (RNN) in reinforcement learning manner to decide the arguments of a convolutional module and hyperparameters like the number of filters~\cite{zoph2016neural}. In NASNet, Zoph et al. proposed the two steps searching strategy that at first it looks for two kinds of `Cell' called basic units based on small scale datasets and then builds up the final solution for larger datasets with piles of these `Cell's~\cite{zoph2018learning}. One major problem with NASNet is the enormous cost, which typically amounts to hundreds of GPUs working together for a few days. How to reduce the consumption on time and memory space and simultaneously maintain a high effectiveness has long been a hot topic in the NAS field. The progressive NAS (PNAS) gains seven times faster speed than NASNet through a smaller searching space, a heuristic searching strategy and an empirical surrogate evaluator~\cite{pnas}. The efficient NAS (ENAS) enormously improves the efficiency of NASNet by means of weights sharing among basic modules in the searching space~\cite{enas}. To further cut the expenditure on memory, researchers turned their eyes on the hypernetwork learning controller. The one-shot NAS~\cite{one_shot} and the proxyless NAS~\cite{proxyless} get rid of the reinforcement or evolutionary learning controller, they arrange each operation unit a probability and train them once for all directly on the target dataset rather than a smaller dataset. The differentiable architecture search (DARTS) introduced a mathematical relaxation from discrete searching to continuous searching and replaced the controller learning with a gradient descent updating process~\cite{darts,pcdarts}. 

The success of NAS in classification tasks also stimulates the endeavors in the semantic segmentation scenarios. Chen et al. demonstrated the feasibility by recursively searching the encoder and decoder blocks~\cite{nas4seg}. Liu et al. devised the Auto-DeepLab and brought up the idea of hierarchical searching space which covers both the `Cell' level and the backbone network level~\cite{auto_deeplab}.
When it comes to medical imaging segmentations, the NAS-Unet respectively looks for the optimal basic down-sampling and up-sampling units and constructs the U-Net shaped architectures, which has been tested on 2D medical image datasets including the prostate MRI, liver computed tomography (CT) and nerves ultrasound images~\cite{nas_unet}. Zhu et al. developed a DARTS-style differentiable NAS U-Net for segmentation of the lung and pancreas on 2D and 3D CT datasets~\cite{vnas}. 

In this paper, we present the NAS based solution for the multi-labeled glioma volumetric segmentation on four modalities of structural MRI scans. Patching strategies cutting a big image into small pieces have been employed due to the high resolution of the input dataset. The searching process takes place on small sized patches, while the training and test processes are undergone on patches with higher contrast. The basic unit to be searched has three different types, correspondingly there are three kinds of searching spaces. NAS is in charge of finding the best building blocks for the downward and upward sampling in the U-Net. Since the input multimodal MRI data are four dimensional matrices and all the deep learning modules work in their 3D versions, throughout this article we call this proposed solution the NAS-3D-U-Net. The multimodal brain tumor image segmentation (BraTS) benchmark~\cite{brats_1,brats_2,brats_3} has been employed as the testbed for our developed algorithms.   

The contributions of this work boils down to threefold:  
\begin{itemize}
\item One NAS based solution for multimodal volumetric MRI gliomas segmentation task has been proposed, which could liberate the network designers from the laborious parameter tuning work in the long run. 
\item The empirical searching strategy of learning two categories of parameters alternately on different datasets has been proved effective for the brain tumor segmentation tasks on the BraTS 2019 Dataset. 
\item Last but not the least, we bring up a brain-wise normalization and a patching strategy specifically for the brain MRI processing. 
\end{itemize}

The remainder of this paper is organized as follows. Section 2 explores details about the searching and training procedures of the NAS-3D-U-Net. Then in section 3 we exhibit and discuss the brain tumor segmentation results. Finally, there is a conclusion on this work in Section 4.

\setcounter{footnote}{0}
\section{Method}
\subsection{Prerequisite}\label{sec_pre}

\begin{figure}[!htbp]
  \centering
\includegraphics[width=0.65\textwidth]{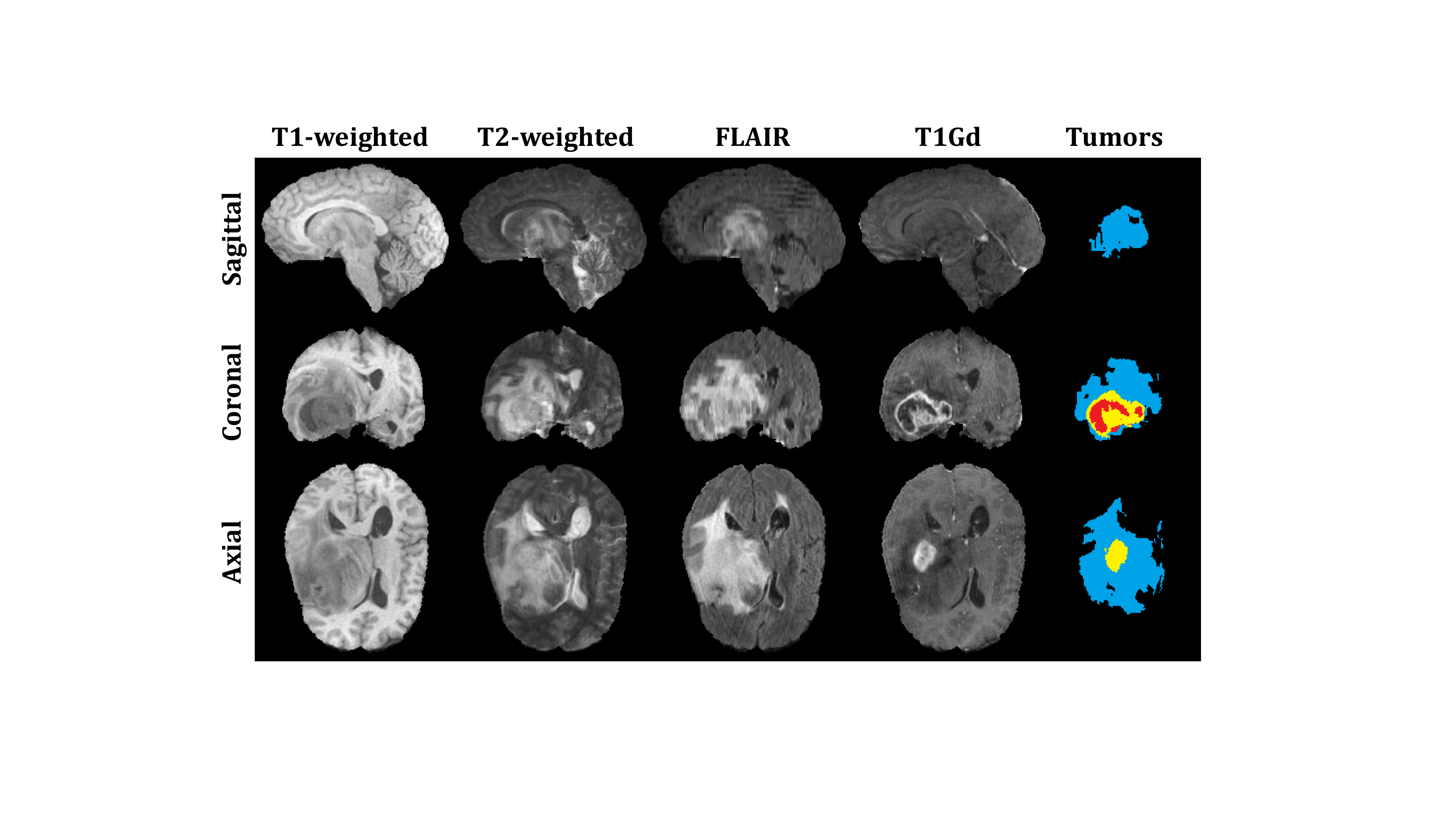}
  \caption{An instance of MRI slices in four modalities and the annotated tumor delineations. `Sagittal', `Coronal' and `Axial' are terminologies indicating the three dimensions. Different colors in the last column represent different levels of tumor tissues: red for the necrotic and non-enhancing tumor core, yellow for the enhancing tumor core, and blue for the edema~\cite{brats_1}. (Bear in mind the colors are just for illustration purposes but not reflecting the real contrast.)}
  \label{fig0}
\end{figure}

For glioma segmentation tasks, all the provided structural MRI pictures are 3D volumes revealing the distribution of hydrogen proton energy in brain tissues. More specifically, at the beginning of the scan, the hydrogen nuclei in human body are spinning in different phases and the axes they rotate on are oriented in the same direction as the magnetic field of the MRI scanner. Then an extra radio frequency (RF) pulse is introduced and the protons are forced to realign to the new orientation and spin in the same phase. When the pulse stops, the protons are coming back to the former state, during which they will emit the absorbed energy. The MRI scanner monitors these emissions and maps them into gray scale images. The speed of proton realignment and phase transition varies for distinct tissues. As a result, by controlling the scanning time intervals we can have different types of tissue to be highlighted. The two tunable variables are the repetition time (TR) which decides the time slot between two RF pulses and the time to echo (TE) which constraints the time span between the RF pulse generation and the emitted signal reception. In ascending order of TR and TE, the three most commonly used MRI modalities are T1-weighted, T2-weighted and the fluid attenuated inversion recovery (FLAIR)\footnote{\scriptsize{T1: TR$\approx$500, TE$\approx$14; T2: TR$\approx$4000, TE$\approx$90; FLAIR: TR$\approx$9000, TE$\approx$114.}}. Constants T1 (the longitudinal relaxation time) and T2 (the transverse relaxation time) reflects the respective time protons need for the realignment and the phase transition. Correspondingly, the contrast and brightness of T1-weighted and T2-weighted images are predominantly determined by the T1 and T2 properties of tissues. FLAIR is pathologically sensitive, it suppresses the free fluid and  lightens the pathological tissues. Besides, in this work we also take into account another T1-weighted MRI with gadolinium (T1Gd). Gadolinium works as an injected contrast agent that helps to enhance the tumor areas. Instances of these four modalities mentioned above could be found in Fig.~\ref{fig0} which also illustrates the ground truth tumor images.

\subsection{Preprocessing}
In this work, the preprocessing takes place under the skull, which means we do a z-score normalization and min-max scaling for each volume without taking the black background into account. Mathematically speaking, for each modality, there is a mean value $\mu$ and standard deviation $\sigma$ of all the non-zero valued voxels in the whole number of training volumes. Given $\bm{A}\in \mathbb{R}^{D_{\rm{S}}\times D_{\rm{C}}\times D_{\rm{A}}}$ represents an original input MRI image, $D_{\rm{S}}$, $D_{\rm{C}}$ and $D_{\rm{A}}$ refer to the volume size on three dimensions. The preprocessing is carried out in the following way: 

\begin{align}
	\hat{A}_{ijk} &=
    \begin{cases}
        (A_{ijk}-\mu)/\sigma & \text{if $A_{ijk}\not=0$} \\
        \qquad\ 0 & \text{else,}
    \end{cases}\label{eq1}\\
    \tilde{A}_{ijk}&=
    \begin{cases}
    	\xi\bigg(\dfrac{\hat{A}_{ijk}-\hat{A}_{\rm{min}}}
        {\hat{A}_{\rm{max}}-\hat{A}_{\rm{min}}}
                         +\lambda\bigg) & \text{if $A_{ijk}\not=0$}\\                         
        \qquad\   0  & \text{else,}  
    \end{cases}\label{eq2}
\end{align}
in which $i\in[1,D_{\rm{S}}]$, $j\in[1,D_{\rm{C}}]$, and $k\in[1,D_{\rm{A}}]$. $\hat{A}_{\rm{min}}$ and $\hat{A}_{\rm{max}}$ indicate the minimum and maximum values for all the $\hat{A}_{ijk}$ whose corresponding $A_{ijk}\not=0$.  $\xi$ and $\lambda$ are two constants used for discriminating the normalized brain voxels from the background. $\tilde{\bm{A}}$ is the preprocessed image.

\setcounter{footnote}{0}
\subsection{Patching Strategy}
\begin{figure}[!htbp]
\centering
\includegraphics[width=0.45\textwidth]{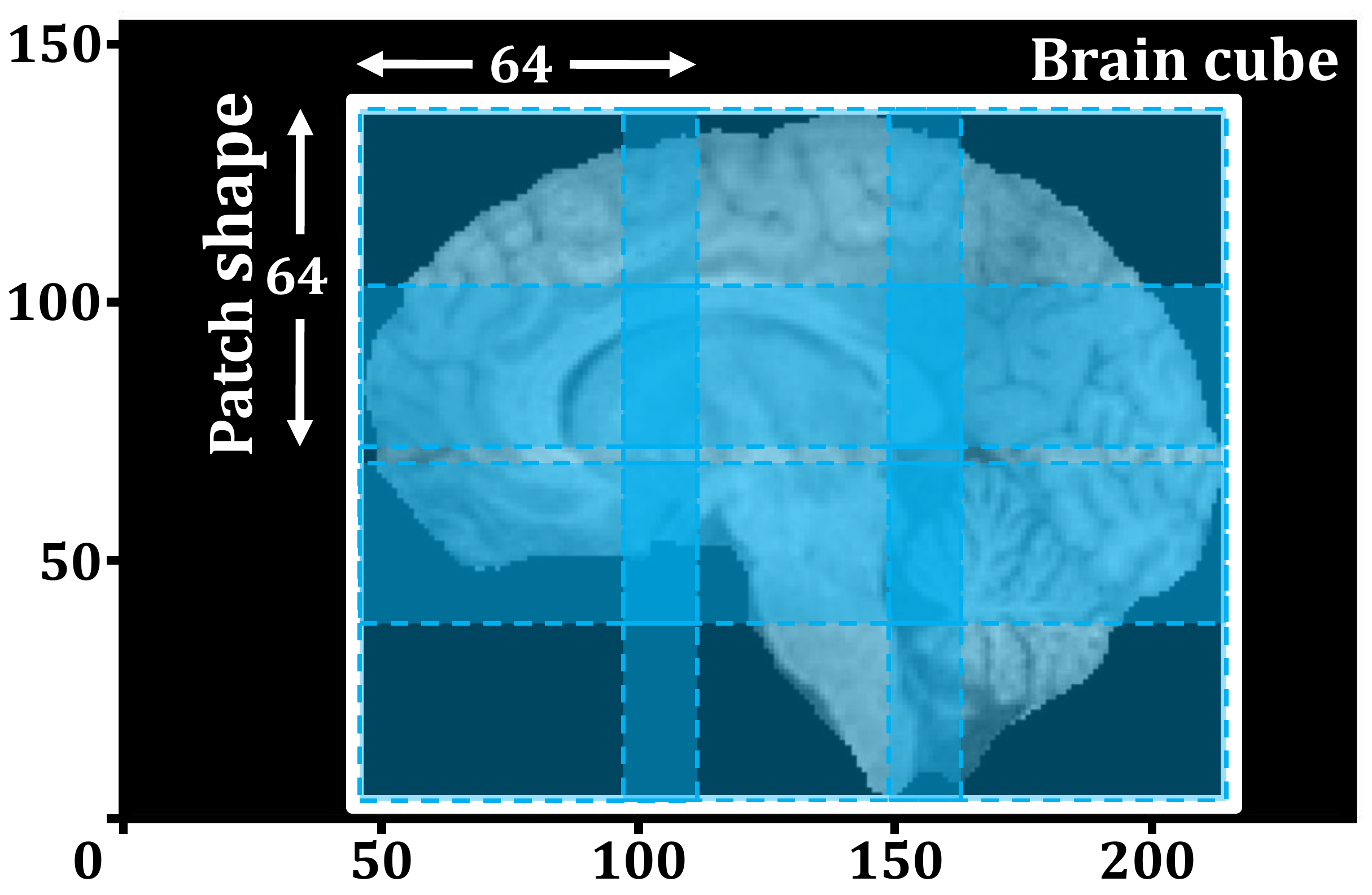}
\caption{An example of the auto-fitting patching strategy deployment. This is a sagittal slice of a T1-weighted MRI volume in size of $240\times240\times155$. The shape of each patch is $64\times64\times64$. The white rectangle indicates the brain cube and the blue dashed line blocks represent the patches. There are nine patches overlapping with each other.} \label{patching}
\end{figure}

Patching strategies would cut a large input image into smaller ones and stitch the corresponding small sized output volumes together. Through this way, it is theoretically possible for a memory limited GPU to deal with pictures in any size. In this work, a patching strategy which we call `auto-fitting' is deployed. The idea of auto-fitting is to cover the brain encapsulated space with the least number of volumetric patches which are symmetrically located. For each MRI image, we define a property named `brain cube' to record the three dimensional scales of the brain area. On each axis, given $l_{\rm{b}}$ and $l_{\rm{p}}$ respectively represent the lengths of the brain cube and the patch, with assumption of $l_{\rm{b}}>l_{\rm{p}}$. Then the number of patches on that axis would be $n_{\rm{p}}=\lceil l_{\rm{b}}/l_{\rm{p}}\rceil$ and the length of the overlap would be $l_{\rm{o}}=\lfloor\frac{n_{\rm{p}}l_{\rm{p}}-l_{\rm{b}}}{n_{\rm{p}}-1}\rfloor$. Suppose the brain cube starts at 0 on that axis, then the starting point of the first patch would be $-\lfloor\frac{n_{\rm{p}}l_{\rm{p}}-l_{\rm{o}}(n_{\rm{p}}-1)-l_{\rm{b}}}{2}\rfloor$ and the moving step of the patches would be $l_{\rm{p}}-l_{\rm{o}}$. Fig.~\ref{patching} illustrates an example of the patching arrangement on a sagittal slice. 

\subsection{Neural Architecture Search}
\subsubsection{Backbone Network: }
\begin{figure}[!htbp]
\centering
\includegraphics[width=0.6\textwidth]{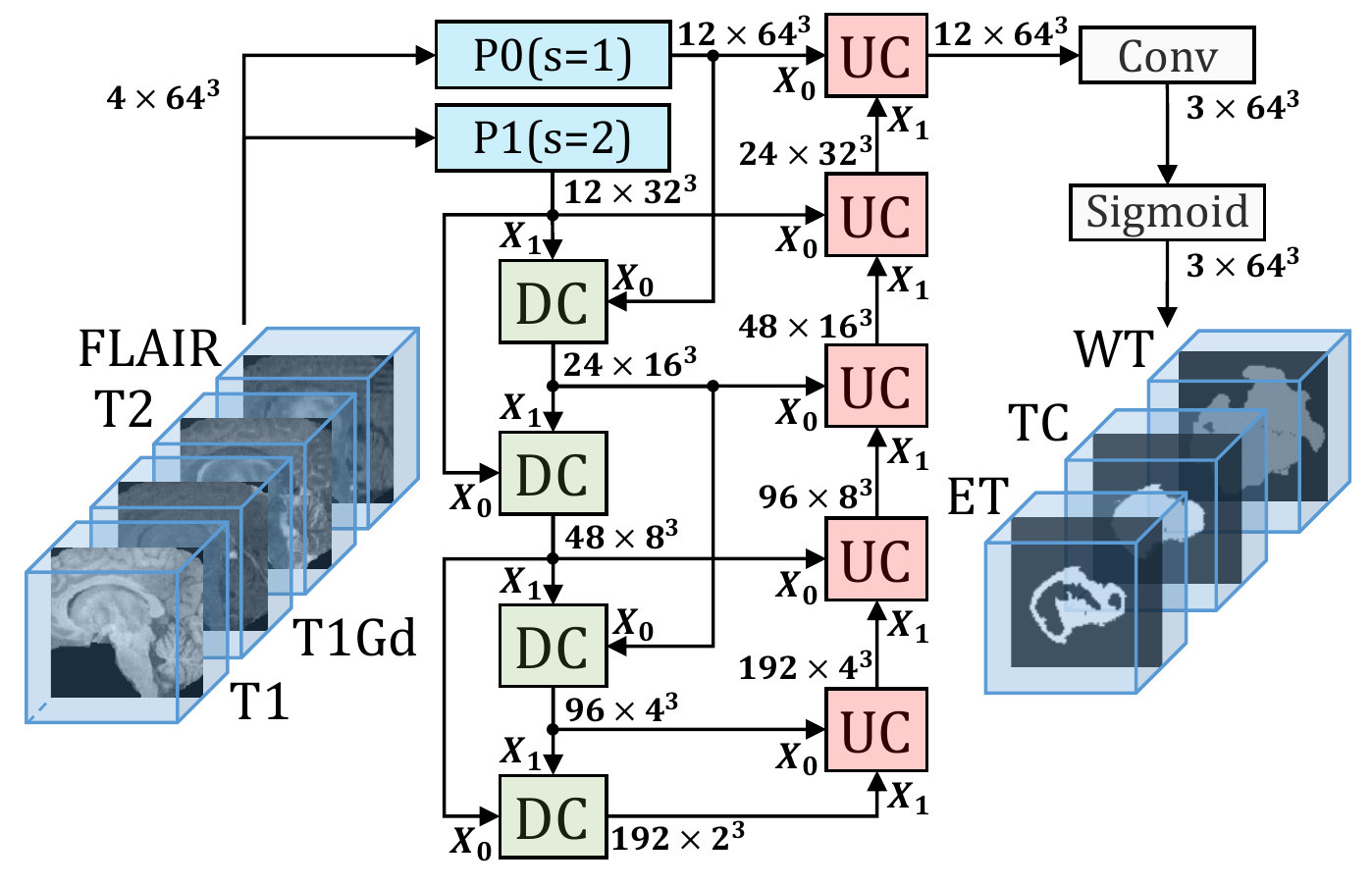}
\caption{Schematic of the NAS-3D-U-Net. This figure shows an example of the three tumor subregions segmentation task on BraTS 2019 dataset. The first channel of the input shape repersents the number of modalities, $64^3$ indicates the patch shape. Both P0 and P1 are composed by a Conv layer and a GN layer, parameter `s' refers to the stride. There are two inputs and one output for each downward Cell (DC) and upward Cell (UC). In accordance with the disciplines in BraTS 2019, herein the three tumor subregions are: 1) the enhancing tumor (ET) which is equal to the enhancing tumor core. 2) the tumor core (TC) including the necrotic and non-enhancing tumor core and ET. 3) the whole tumor (WT) which contains TC and the edema.} \label{schem}
\end{figure}

The essence of U-Net is the U-shaped architecture composed of mutually connected down sampling and up sampling blocks. In NAS-3D-U-Net, the macro structure still consists of one downward route and one upward route. Nonetheless, when it comes to the micro structure of each individual block on these routes, the NAS would be in charge of the organization management. Following the convention of NAS, the generated building blocks for the outside network are called `Cell's. Throughout this work, the down-sampling and up-sampling blocks are named downward Cell (DC) and upward Cell (UC) respectively. 
The DCs take responsibility for the feature embedding, which compresses the resolution and extracts the target sensitive information. The UCs would mix the embedded features with former reserved DC outputs which stores the inevitably lost positional information during the down-sampling. The constitution of NAS-3D-U-Net has been depicted in Fig.~\ref{schem} which takes the BraTS dataset for instance. Ahead of DCs there are two primary modules P0 and P1 which are made up of one 3D convolutional (Conv) layer followed by a 3D group normalization (GN) layer (We choose GN in this work because the batch size is very small~\cite{group_norm}). P0 doesn't change the image contrast while P1 shrinks the image into its half size. The number of kernels (or filters) in P0 and P1 are decided according to the number of nodes, which is the basic operation union in a Cell and will be explored in the next part. Right after the last UC, the contrast recovered data is transformed into the degree of confidence for three types of labels through the Conv and sigmoid layers.  

\subsubsection{Searching Space: }
\begin{figure}[!htbp]
\centering
\includegraphics[width=0.3\textwidth]{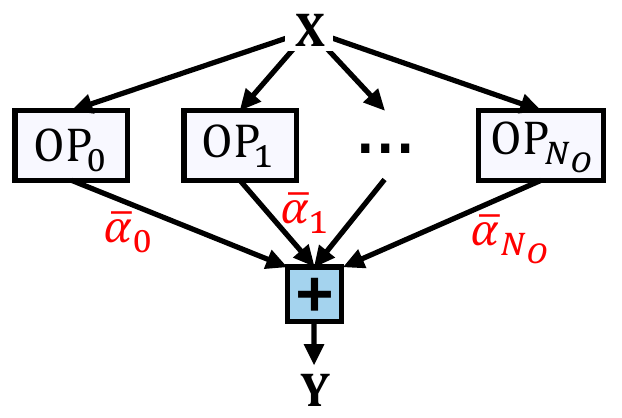}
\caption{Hybrid module structure. ${\rm{OP}}_i$ refers to the individual operation and $\bar{\alpha}_i$ is the corresponding weight parameter, $i\in[1,N_{\rm{O}}]$.} \label{hm}
\end{figure}

The explanation of the searching space starts with the definition of a hybrid module (HM) which is the fundamental computing unit in DC and UC. As illustrated in Fig.~\ref{hm}, the HM is a mixture of $N_{\rm{O}}$ different operations (OPs), with an assurance that all of the OPs have the same output shape. For each ${\rm{OP}}_i$, $i\in[1,N_{\rm{O}}]$, there is a parameter $\alpha_i$ whose softmax transformation $\bar{\alpha}_i={\rm{exp}}(\alpha_i)/\sum_{j=1}^{N_{\rm{O}}}{\rm{exp}}(\alpha_j)$ is assigned to the output of ${\rm{OP}}_i$ as a weight. The $\bar{\alpha}_i$ plays as the probability indicating how much this ${\rm{OP}}_i$ contributes to the HM. The output of HM is a weighted sum of all the OPs. As the searching process goes, the optimizers would increase the $\alpha$ whose bonded OPs affect more to the HM while decrease the other $\alpha$s which belong to less important OPs. In the rest of this work, for simplicity, we call $\alpha$ the hybrid parameter (HP) and the rest parameters are referred as the kernel parameter (KP).  

Inspired by the searching space for 2D medical image segmentation tasks~\cite{nas_unet}, in this work we propose three kinds of HMs, including the down sampling HM (DHM), the up sampling HM (UHM) and the normal HM (NHM). The operation sets for each HM have been listed in Table~\ref{tab_hm} from which we can see that four types of Conv modules are picked out and commonly used. 
\begin{table}[!htbp]
\centering
\caption{Operation candidates for three kinds of hybrid modules (HM).}\label{tab_hm}
\begin{tabular}{cllllll}
\hline
\specialrule{0em}{1.5pt}{1.5pt}
HM type & \multicolumn{6}{c}{Operation candidates} \\
\hline
\specialrule{0em}{1.5pt}{1.5pt}
DHM &\   d\_conv \quad& d\_dil\_conv \quad& d\_dep\_conv \quad& d\_se\_conv \quad& max\_pool \quad& avg\_pool\\
UHM &\  u\_conv & u\_dil\_conv & u\_dep\_conv & u\_se\_conv\\
NHM &\  conv & dil\_conv & dep\_conv & se\_conv & identity\\
\hline
\end{tabular}
\end{table}
The `conv' represents the basic $3\times3\times3$ Conv layer. The `dil\_conv' refers to the $3\times3\times3$ dilated convolution which has an enlarged reception field~\cite{dil_conv}. The `dep\_conv' is the depthwise separable convolution in which the function of a $3\times3\times3\times4$ shaped Conv kernel could be implemented by a combination of four $3\times3\times3\times1$ depthwise Conv kernels and one $1\times1\times1\times4$ pointwise Conv kernel~\cite{dw_conv}. The `se\_conv' is short for the squeeze and excitation convolution which brings the attention mechanism in by means of learning the significance distribution among different channels~\cite{se_conv}. The prefix `d\_' in a DHM indicates the stride is two for that Conv layer which would change the resolution into half size. On the contrary, the Conv operations with prefix `u\_' in UHM are transposed convolutions~\cite{transposed_conv} which will have the image contrast doubled. All these Conv OPs involve the GN and ReLU activation layers coming after the convolutions. Aside from the Conv modules, DHM also has the max pooling layer (`max\_pool') and average pooling layer (`avg\_pool'). NHM has the `identity' OP which only does the GN and ReLU calculations. Last but not the least, none of the HMs would change the channel size.

\begin{figure}[!htbp]
  \centering
  \subfigure[]{
    \label{dc}
    \includegraphics[width=0.35\textwidth]{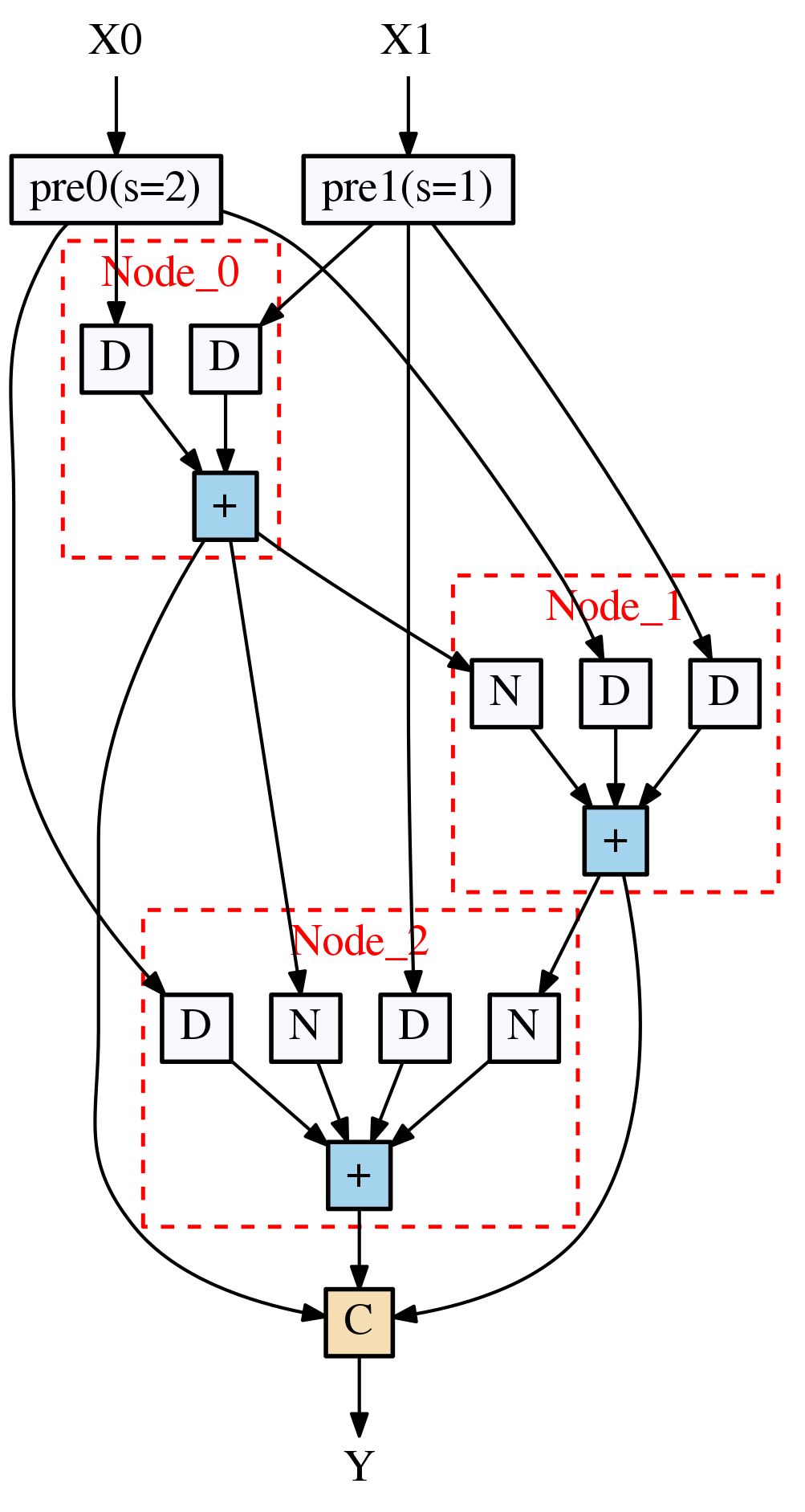}}
  \hspace{0.4in}
  \subfigure[]{
    \label{uc}
    \includegraphics[width=0.35\textwidth]{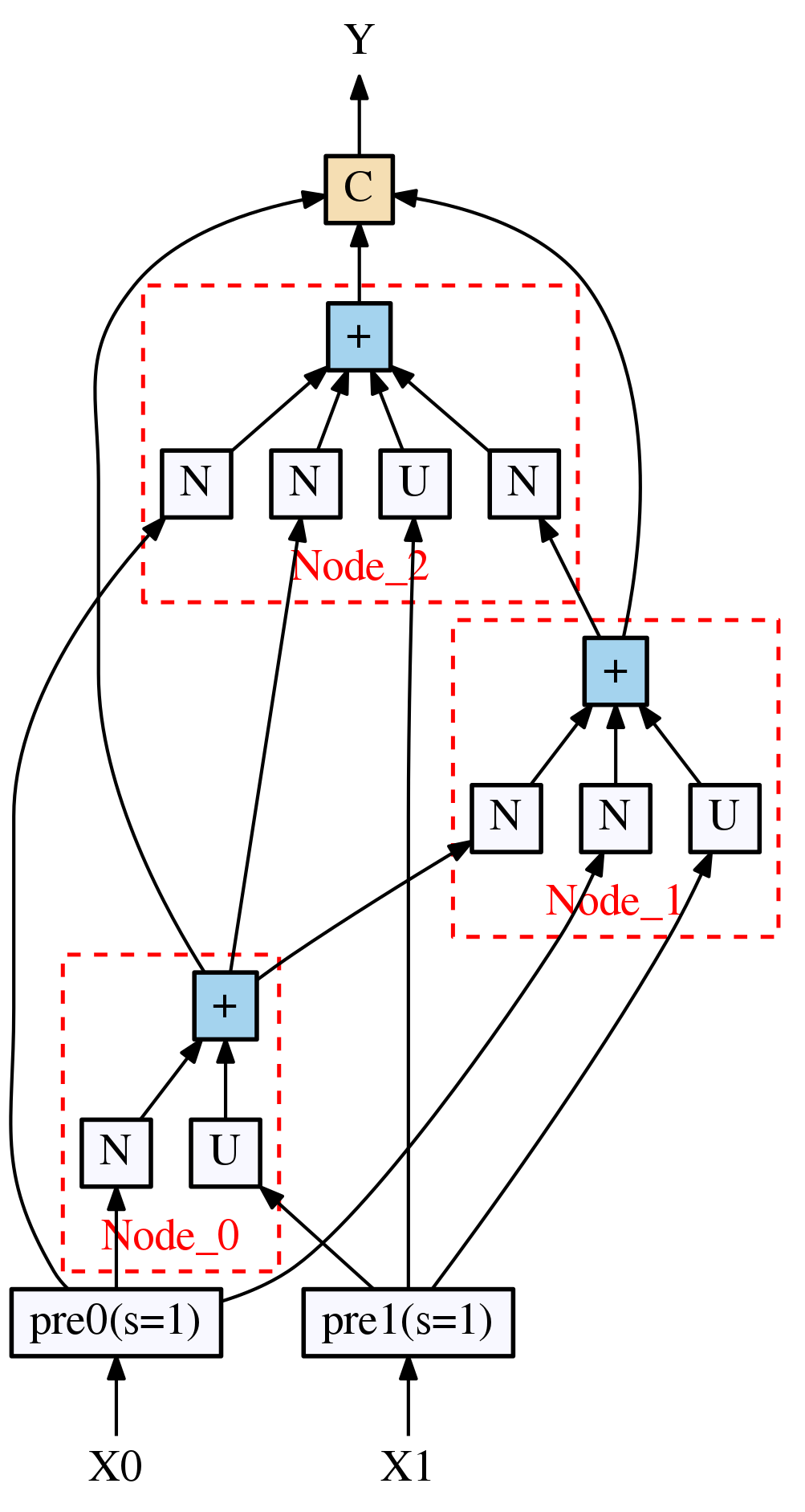}}
  \caption{Diagrams of Cells. (\textbf{a}) The downward Cell (DC). (\textbf{b}) The upward Cell (UC). The blocks with `pre0' and `pre1' represent the preprocessing Conv layers with certain stride arguments. `D', `U' and `N' are respectively short for the downward, upward and normal hybrid modules (HM). The plus signs in blue squares indicate element-wise additions and the yellow squared `C' is the concatenation on the channel dimension. The conception of the Node has been illustrated with red dashed lines and labels.}
  \label{cells}
\end{figure}

With hybrid modules at hand, we are able to assemble the DC and UC as exhibited in Fig.~\ref{cells}. Here we define a `Node' as a cluster of HMs encircled by the red dashed lines. In this work, each Cell has three Nodes, the number of HMs in these Nodes are in ascending order as the output of the previous Node would be taken as the input signal for the next Node. Finally, the output of a Cell is a concatenation of three nodes' outputs each of which is an accumulation of all the HMs in the certain node.       

According to the design of NAS-3D-U-Net displayed in Fig.~\ref{schem}, the two inputs of a Cell always have different shapes. Hence in Fig.~\ref{cells}, two preprocessing maneuvers which are mainly $1\times1\times1$ Conv layers have been set up before the HMs. Assuming the MRI dataset has $m$ modalities (channels) and there are $n$ Nodes in each Cell whose output channel we want to be $\theta$ times as large or small as the input channel, then the number of kernels for P0 and P1 (in Fig.~\ref{schem}) would be set to $m\cdot n$ and the output channel of the $i$th DC or UC (from top to bottom in Fig.~\ref{schem}) would be $m\cdot\theta^{i}$, in which $i\in \mathbb{N}^+$ and $\theta$ is called the zoom factor.

\subsubsection{Searching Strategy: }
\begin{figure}[!htbp]
  \centering
  \subfigure[]{
    \label{searched_dc}
    \includegraphics[width=0.38\textwidth]{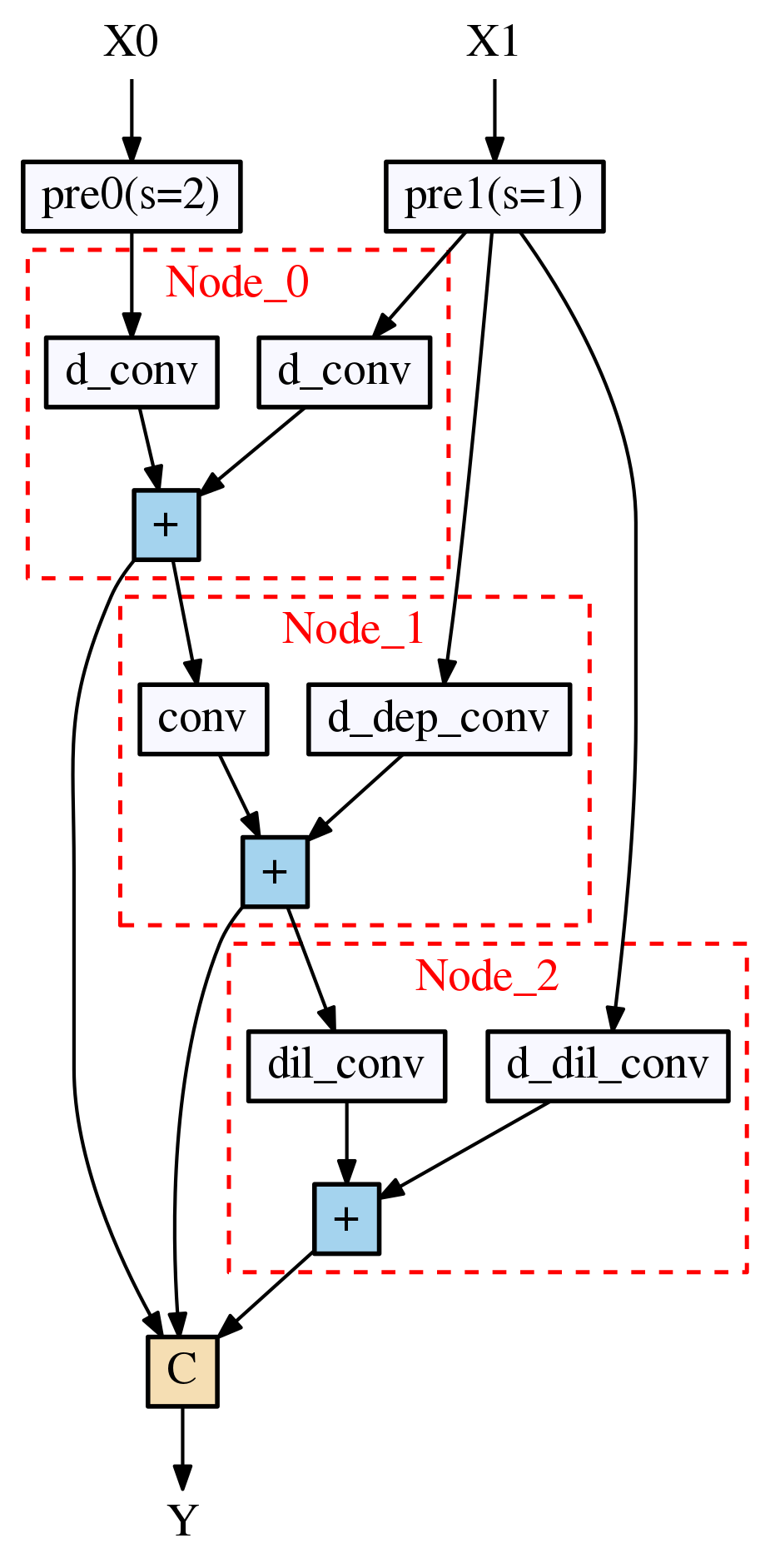}}
  \hspace{0.3in}
  \subfigure[]{
    \label{searched_uc}
    \includegraphics[width=0.5\textwidth]{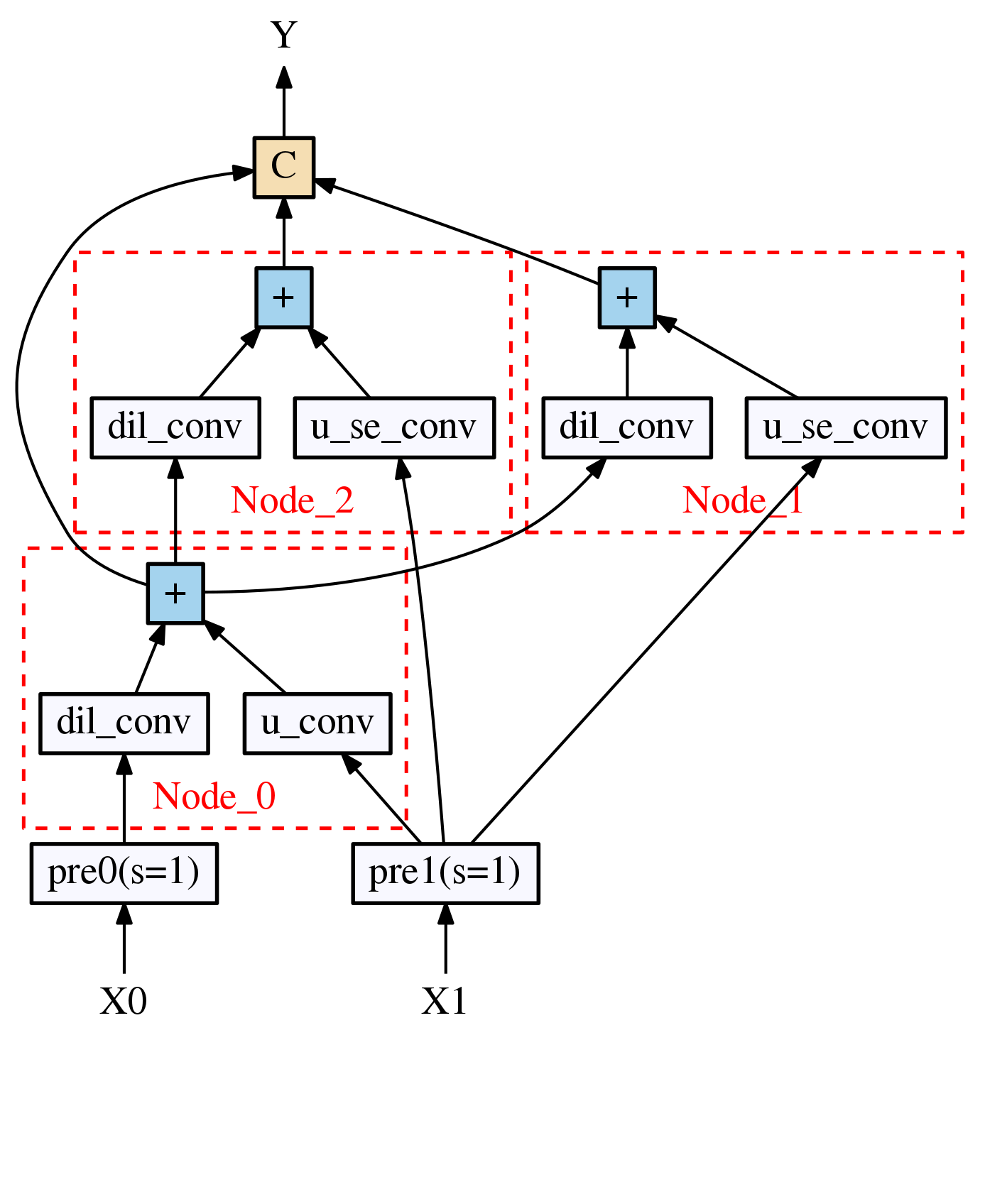}}
  \caption{The NAS-3D-U-Net generated Cells (\textbf{a}) The searched downward Cell. (\textbf{b}) The searched upward Cell. The rectangles with operation names (which can be found in Table~\ref{tab_hm}) represent the selected hybrid modules in the Nodes.}
  \label{searched_cells}
\end{figure}

The searching process is to learn the $\alpha$ parameters in HMs and decide the structure of DC and UC. Given a set of $\alpha$, each HM in a Cell would only keep one OP with the highest $\bar{\alpha}$ value, and in each Node only the first two highly ranked HMs would be finally elected. Fig.~\ref{searched_cells} displays the searched architectures of DC and UC on the BraTS 2019 dataset, which gives an example of how does the generated Cells look like.  

In this work, we carry out an empirically searching strategy that alternately updates the hybrid parameters and the kernel parameters in the network, the detail of which has been concluded in Table~\ref{algorithm}. For each category of parameters there is going to be one optimizer, and the two optimizers would work sequentially in each iteration. The data for searching are separated into two parts, one for hybrid parameters learning and the rest for the kernel parameters learning. Correspondingly, the respective loss values calculated on these two sets are referred to as the hybrid loss and the kernel loss. Once the best Cell structures are found, we will replace the DCs and UCs with the searched architectures and retrain the network on the same dataset. 

\begin{table}[!htbp]
\centering
\caption{Searching strategy in NAS-3D-U-Net.}\label{algorithm}
\begin{tabular}{cl}
\hline
\specialrule{0em}{1.5pt}{1.5pt}
\multicolumn{2}{l}{Searching algorithm:} \\
1 & Prepare the datasets for hybrid parameters and kernel parameters. \\
  & Let $\{C\}$ record the searched Cell structures and their counts.\\
  & $N_C:=$ the minimum number of counts we need to find a best Cell $\hat{C}$.\\
  & $N_E:=$ the total number of epochs. $\hat{C}:=0$.\\
2 & \textbf{for} $i$ \textbf{in} 1 \textbf{to} $N_E$:\\   
3 & \quad Get the searched Cell structure $C_i$.\\
4 & \quad The count of $C_i$ in $\{C\}$ += 1.\\
5 & \quad \textbf{if} the count of $C_i\ge N_C$:\\
6 & \qquad $\hat{C}=C_i$.\\
7 & \qquad \textbf{break}.\\
8 & \quad Get the hybrid loss; back propagation; update the hybrid parameters.\\
9 & \quad Get the kernel loss; back propagation; update the kernel parameters.\\
10 & \textbf{if} $\hat{C} \ne 0$:\\
11 & \quad $\hat{C}$ = the most common Cell structure in $\{C\}$.\\
12 & \textbf{return} $\hat{C}$.\\
\specialrule{0em}{1.5pt}{1.5pt}
\hline
\end{tabular}
\end{table}

\section{Experiment and Results}
\subsection{Dataset}

The BraTS 2019 dataset we used in this work is coming from the multimodal brain tumor segmentation challenge 2019 which is an annually hosted contest since 2012~\cite{brats_1}. Aiming at pushing the advance of computer vision solutions for brain tumor diagnosis and treatments, BraTS keeps providing abundant clinically acquired MRI scans~\cite{brats_2,brats_3}. The BraTS 2019 dataset has collected pre-operative multimodal MRI scans and the neuroradiologists verified tumor labels of subjects with the glioblastoma/high grade glioma (HGG) or the low grade glioma (LGG) from 19 institutions. The four MRI modalities are T1-weighted, T2-weighted, FLAIR and T1Gd as we described in section~\ref{sec_pre}. All the MRI and label volumes are in shape of $240\times 240\times 155$.   
The ground truth labels which are pathologically confirmed by experts have four voxel values: 1 representing the necrotic and non-enhancing tumor core, 2 referring to the edema, 4 indicating the enhancing tumor core, and 0 covering the other places. The tumor subregions considered in the evaluation system are inclusive combinations of value 1,2 and 4. Specifically, the enhancing tumor (ET) is 4, the tumor core (TC) includes 1 and 4, the whole tumor (WT) equals to the complete set of 1, 2 and 4. Samples of the dataset could be seen in Fig.~\ref{fig0}. 
The BraTS 2019 Training Dataset is made up by MRI scans and ground truth labels of 259 HGG and 76 LGG subjects. The BraTS 2019 Validation Dataset for generalization and scalability verification only provides the MRI volumes of 125 subjects, which in this work are leveraged as the testing dataset. All of the MRI data have undergone preprocessing including the registration, $1\times1\times1$ millimeter resolution resampling, and skull stripping. 
\subsection{Loss Function}
The loss function integrated in NAS-3D-U-Net is the weighted multi-class Dice loss which has been demonstrated as an effective variant of Dice loss for brain tumor segmentation tasks~\cite{isensee2017brain,vnet}. As exhibited in Eq.~(\ref{eq3}),
\begin{equation}
    L=1-\frac{1}{3}\sum_{h=1}^{3}\frac{\epsilon+2\sum\limits_{i=1}^{P_{\rm{S}}}\sum\limits_{j=1}^{P_{\rm{C}}}\sum\limits_{k=1}^{P_{\rm{A}}}Y_{hijk}\hat{Y}_{hijk}}
            {\epsilon+\sum\limits_{i=1}^{P_{\rm{S}}}\sum\limits_{j=1}^{P_{\rm{C}}}\sum\limits_{k=1}^{P_{\rm{A}}}Y_{hijk}+\sum\limits_{i=1}^{P_{\rm{S}}}\sum\limits_{j=1}^{P_{\rm{C}}}\sum\limits_{k=1}^{P_{\rm{A}}}\hat{Y}_{hijk}},
\label{eq3}
\end{equation} 
in which $P_{\rm{S}}$, $P_{\rm{C}}$ and $P_{\rm{A}}$ indicate the patch shape on three dimensions, $\epsilon$ is a tiny constant to avoid zero division error. $\bm{Y}$ indicates the $3\times P_{\rm{S}}\times P_{\rm{C}}\times P_{\rm{A}}$ sized matrix extracted from the ground truth label, and $\hat{\bm{Y}}$ represents the predicted output.

\subsection{Configurations}
The NAS-3D-U-Net has been developed with a single GTX1080Ti GPU card and PyTorch framework. For brain-wise normalization in Eq.~(\ref{eq2}), we set $\xi=100$ and $\lambda=0.1$. The patch size is $64\times64\times64$ for the searching process and $128\times128\times128$ for the training process. The hybrid parameter $\alpha$ is initialized as 0. In a Cell, the number of Nodes $n=3$ and the zoom factor $\theta=2$. For loss function in Eq.~(\ref{eq3}) $\epsilon=1e-6$. In the data stream pipeline, data augmentations including random distortion, flipping, and rotation are implemented on the fly. The BraTS 2019 Training Dataset are split in the 5-folds cross validation style. The batch size is 1 for all the scenarios in searching and training processes. When stitching the patches, we keep the average as the value for the overlapped voxel. 

\subsection{Searching Results}
\begin{figure}[!htbp]
\centering
\includegraphics[width=0.4\textwidth]{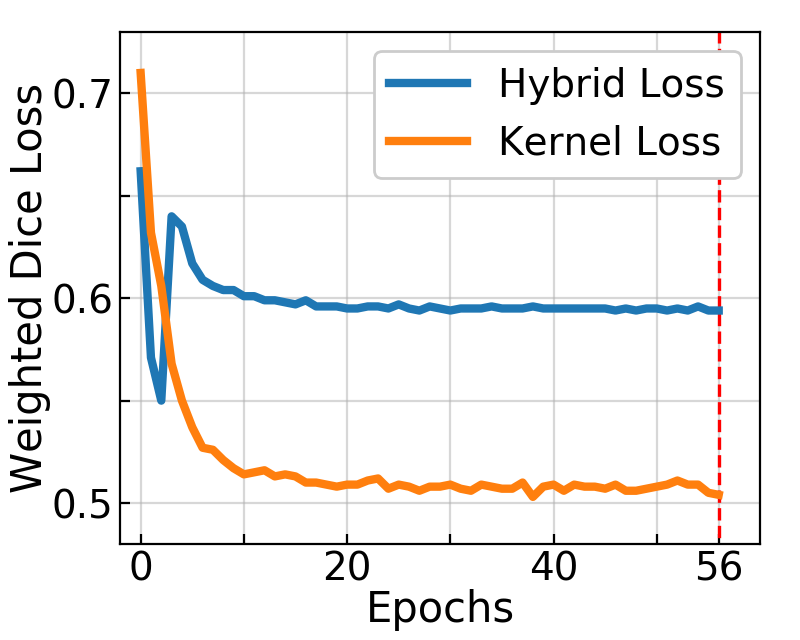}
\caption{Searching history. Notice the learning processes for both hybrid and kernel parameters are applied alternately in the same epoch. The last epoch has been highlighted by the red dashed line.} \label{searching_log}
\end{figure}

For the searching process, we set $N_E=40$, $N_C=100$ and recored the searched Cell structures in a hash-map. One fifth of the training set are used for updating hybrid parameters and the others are for the kernel parameters. Histories of the hybrid loss and the kernel loss have been shown in Fig.~\ref{searching_log}, from which we can see the hybrid loss value vibrated a lot during the first 10 epochs and then converged to a stable state. The iteration was stopped by the break sentence after the 56th epoch when the searched structure $\hat{C}$ appeared for its 40th time. 

The searched Cell structures have been depicted in Fig.~\ref{searched_cells}. One obvious characteristic shared by the searched DC and UC is that X1 is fed into every Node while X0 only affects the first Node. This phenomenon partially attributes to the fact that for both DC and UC, X1 is directly coming from the previous Cell whereas X0 is a shortcut from even further Cells. Another detail we have noticed is that in DC the resolution of all inputs would be compressed in the first place, which we believe would create more diversity for the next layer. In Fig.~\ref{searched_uc}, the positional information to be recovered is carried by X0, and we can see that information has been merged with the embedded features from X1 in the first Node and then propagated to the other two Nodes and the final output signal. 

\subsection{Training and Validation Results}

\begin{figure}[!htbp]
  \centering
\includegraphics[width=0.55\textwidth]{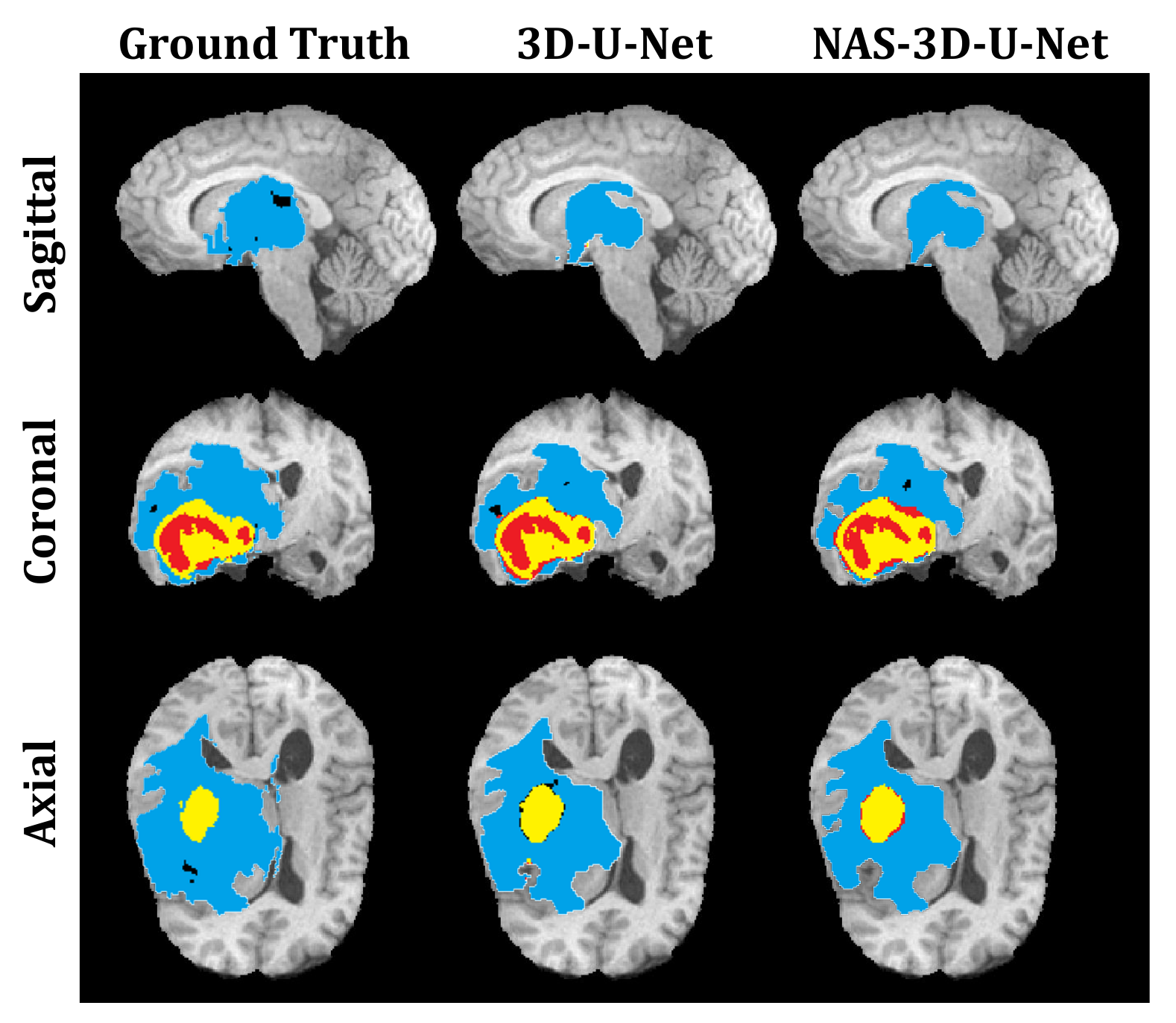}
  \caption{An example of the segmentation results by 3D-U-Net and NAS-3D-U-Net. For illustration purpose, the tumor image has been overlaid on the T1-wighted slices. The colors for tumor subregions have the same meaning as in Fig.~\ref{fig0} beforehand.}
  \label{result}
\end{figure}

In order to prove the feasibility and the scalability of the NAS-3D-U-Net, in this part, we compare the proposed solution with the 3D-U-Net which is a manually designed architecture for brain tumor segmentations~\cite{wang2019brain}. The development environment, preprocessing, patching strategies, data augmentation and the configurations for the baseline algorithm are mostly identical to that of the NAS-3D-U-Net. For training process, we deploy the 5-fold cross validation at first and then take use of the whole training dataset again. The last training had 200 epochs.
Fig.~\ref{result} illustrates one sample of the detected tumor subregions by the two methods as well as the accompanied ground truth labels. It is seen that the NAS-3D-U-Net has detected most part of the three tumor subregions as well as what 3D-U-Net could figure out.

Following the BraTS benchmark convention, four metrics are used to evaluate the performances of an algorithm~\cite{brain_0}. For each tumor subregion (ET, TC and WT), let $\bm{T}$ and $\hat{\bm{T}}\in\{0,1\}^{240\times240\times155}$ be the the volumes respectively extracted from the ground truth label and the network output, then the Dice score would be
\begin{equation}
    Dice=\frac{2\sum_{i=1}^{240}\sum_{j=1}^{240}\sum_{k=1}^{155}T_{ijk}\hat{T}_{ijk}}
            {\sum_{i=1}^{240}\sum_{j=1}^{240}\sum_{k=1}^{155}T_{ijk}+\sum_{i=1}^{240}\sum_{j=1}^{240}\sum_{k=1}^{155}\hat{T}_{ijk}}.
\label{dice_score}
\end{equation} 
The Sensitivity (or Recall) reflects the true positive rate which is
\begin{equation}
    Sens=\frac{\sum_{i=1}^{240}\sum_{j=1}^{240}\sum_{k=1}^{155}T_{ijk}\hat{T}_{ijk}}
            {\sum_{i=1}^{240}\sum_{j=1}^{240}\sum_{k=1}^{155}T_{ijk}}.
\label{sens}
\end{equation} 
The Specificity is the complementary set of the false positive rate which equals
\begin{equation}
    Spec=\frac{\sum_{i=1}^{240}\sum_{j=1}^{240}\sum_{k=1}^{155}(1-T_{ijk})(1-\hat{T}_{ijk})}
            {\sum_{i=1}^{240}\sum_{j=1}^{240}\sum_{k=1}^{155}(1-T_{ijk})}.
\label{spec}
\end{equation} 
Besides the three volumetric similarities, the 95\% Hausdoff distance measuring the difference between the tumor boundaries has also been leveraged. Given $\partial\bm{T}$ and $\partial\hat{\bm{T}}$ representing the gradients of $\bm{T}$ and $\hat{\bm{T}}$, the surface of the tumor area could be expressed as $\bm{\mathcal{S}}=\{(i,j,k)|\partial T_{ijk}\neq 0\}$ and $\hat{\bm{\mathcal{S}}}=\{(i,j,k)|\partial \hat{T}_{ijk}\neq 0\}$. Then we have
\begin{equation}
    Haus=\max\{
    \sup_{s\in{\bm{\mathcal{S}}}}^{\sim}
    \inf_{\hat{s}\in\hat{\bm{\mathcal{S}}}}
    d(s,\hat{s}), 
    \sup_{\hat{s}\in{\hat{\bm{\mathcal{S}}}}}^{\sim}
    \inf_{s\in\bm{\mathcal{S}}}
    d(\hat{s},s)
\},
\label{haus95}
\end{equation} 
in which $d(\cdot)$ is the Euclidean distance, $\sup\limits^{\sim}$ means finding the 95\% quantile values rather than the maximum.

\begin{figure}[!htbp]
\centering
\includegraphics[width=1\textwidth]{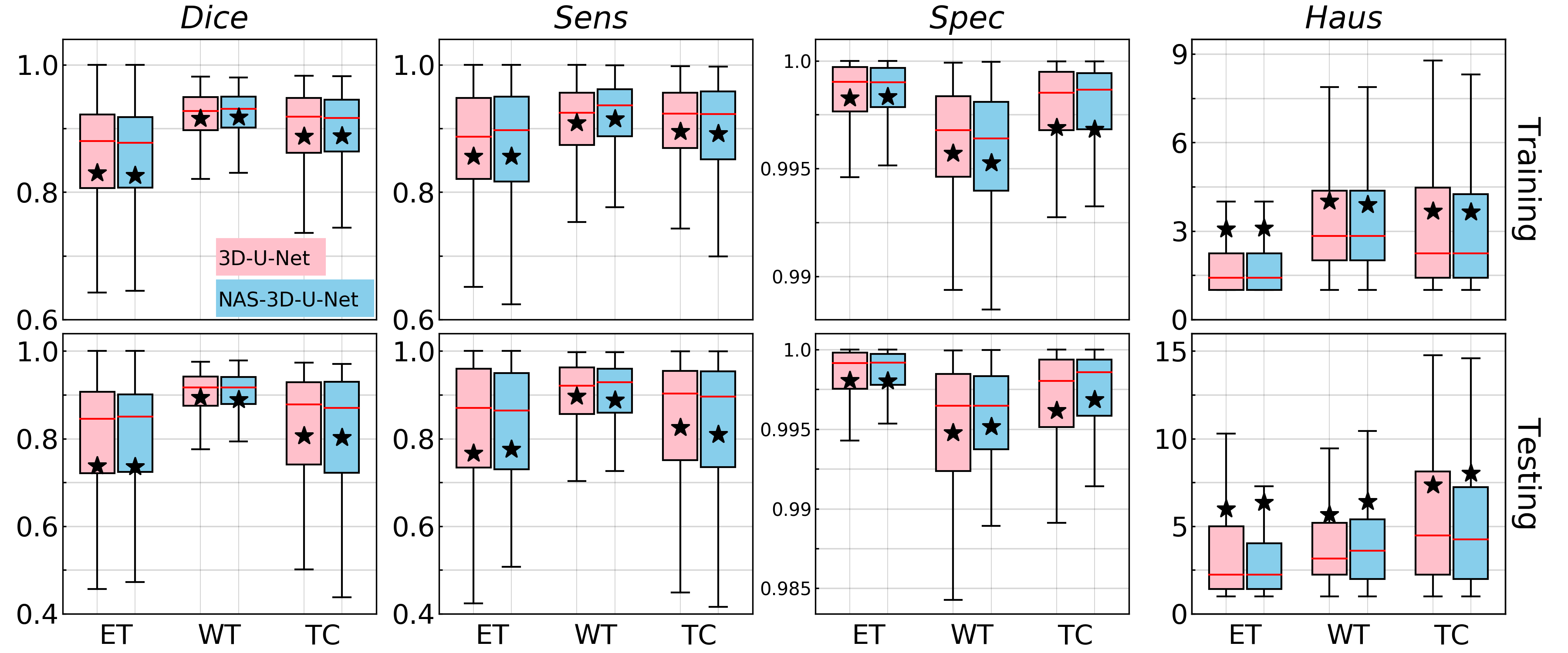}
\caption{Evaluation results of the 3D-U-Net (pink colored) and the NAS-3D-U-Net (blue colored). The first row indicates the results on the training dataset and the second row is for the testing set. Each colum represents one metric evaluation for three tumor subrigions. The median (red line), 25\% quantile, 75\% quantile have been included in the box acommpanied by the 1.5 inter-quartile-ranged whiskers and the mean values (black stars).  } \label{metric}
\end{figure}

In Fig.~\ref{metric} we can see the comparative experiment results in terms of these four metrics. The box plots have shown the distribution characteristics.
From the Dice score and Sensitivity values we can see it is more difficult to correctly predict the TC and ET than that of the WT. On the other hand, the WT doesn't have the smallest Hausdorff distance, which means it is not easy to anticipate the boundary of the WT. This contradictory outcomes reflect the fact that compared with the tumor core the edema may have more irregular surface, which could also be revealed partially from Fig.~\ref{result}.  
In contrast with the manually fabricated 3D-U-Net, the NAS-3D-U-Net shows competitive performances. For most testing cases in Fig.~\ref{metric} the NAS-3D-U-Net has smaller inter quartile ranges. The exceptions happen to the dice score and sensitivity of the tumor core and the hausdoff distance of the whole tumor. 
The accurate mean values for training and testing datasets have also been listed in Table.~\ref{means_train} and Table.~\ref{means_test} respectively. 
From the mean values we could also find that the NAS-3D-U-Net works as well as the 3D-U-Net especially on the testing dataset.

\begin{table}[!htbp]
\centering
\caption{Mean values in evaluations on the training dataset.}\label{means_train}
\begin{tabular}{ccccccccccccccccc}
\hline
\specialrule{0em}{1.5pt}{1.5pt}
\multirow{2}{*}{Algorithm} &\ & \multicolumn{3}{c}{\textit{Dice}}&\ 
						 & \multicolumn{3}{c}{\textit{Sens}}&\ 
						 & \multicolumn{3}{c}{\textit{Spec}}&\ 
						 & \multicolumn{3}{c}{\textit{Haus}}\\
						 &\ &ET&WT&TC&\ &ET&WT&TC&\ &ET&WT&TC&\ &ET&WT&TC\\\hline
						 \specialrule{0em}{1.5pt}{1.5pt}
3D-U-Net &\ &0.83&0.92&0.89&\ &0.86&0.91&0.89&\ &1.00&1.00&1.00&\ &3.07&4.01&3.67\\
\specialrule{0em}{1.5pt}{1.5pt}
NAS-3D-U-Net &\ &0.83&0.92&0.89&\ &0.86&0.91&0.89&\ &1.00&1.00&1.00&\ &3.10&3.90&3.64\\
\hline
\end{tabular}
\end{table}

All these comparisons have demonstrated that the automatically generated network is potentially good enough in both feasibility and scalability to replace the manually designed network for brain tumor segmentation tasks.

\begin{table}[!htbp]
\centering
\caption{Mean values in evaluations on the testing dataset.}\label{means_test}
\begin{tabular}{ccccccccccccccccc}
\hline
\specialrule{0em}{1.5pt}{1.5pt}
\multirow{2}{*}{Algorithm} &\ & \multicolumn{3}{c}{\textit{Dice}}&\ 
						 & \multicolumn{3}{c}{\textit{Sens}}&\ 
						 & \multicolumn{3}{c}{\textit{Spec}}&\ 
						 & \multicolumn{3}{c}{\textit{Haus}}\\
						 &\ &ET&WT&TC&\ &ET&WT&TC&\ &ET&WT&TC&\ &ET&WT&TC\\\hline
						 \specialrule{0em}{1.5pt}{1.5pt}
3D-U-Net &\ &0.74&0.89&0.81&\ &0.77&0.90&0.83&\ &1.00&0.99&1.00&\ &5.99&5.68&7.36\\
\specialrule{0em}{1.5pt}{1.5pt}
NAS-3D-U-Net &\ &0.74&0.89&0.80&\ &0.78&0.89&0.81&\ &1.00&1.00&1.00&\ &6.36&6.40&8.02\\
\hline
\end{tabular}
\end{table}

\section{Conclusion}
In this work, we develop an automated machine learning solution named NAS-3D-U-Net for the 3D multimodal MRI brain tumor segmentation task. Through alternately updating the two classes of parameters the searching process would end up with the most frequently appeared cell structures which are further used as the building blocks in the U-Net architectures. In order to feeding the large 4D input into our networks, the z-score normalization and scaling have been employed only in the brain area. Moreover, NAS-3D-U-Net could take advantage from the patching strategy considering the different patch sizes for searching and training processes. On BraTS 2019 dataset, it has been demonstrated that the autoML searched network has competitive performance in both of the feasibility and generalization.

%
%

\bibliographystyle{IEEEtran}
\bibliography{refs}

\end{document}